\documentclass[prb,showpacs,twocolumn]{revtex4}
\usepackage{mathrsfs}
\usepackage{stmaryrd}
\usepackage{times}
\usepackage{amsmath}
\usepackage{amsfonts}
\usepackage{amssymb}
\usepackage{graphicx}
\usepackage{bm}
\begin{document}

\title{Enhancement of the
thermoelectric figure of merit in a quantum dot due to the Coulomb
blockade effect }
\author{Jie Liu$^{1}$ }
\author{Qing-feng Sun$^{1}$ }
\author{X. C. Xie$^{2,1}$}
\affiliation{$^{1}$Beijing National Laboratory for Condensed Matter
Physics and Institute of Physics, Chinese Academy of Sciences,
Beijing 100190, China} \affiliation{$^{2}$Department of Physics,
Oklahoma State University, Stillwater, Oklahoma 74078, USA}

\begin{abstract}
We investigate the figure of merit of a quantum dot (QD) in the
Coulomb blockade regime. It is found that the figure of merit $ZT$
may be quite high if only single energy level in the QD is
considered. On the other hand, with two or multi energy levels in
the QD and without the Coulomb interaction, the $ZT$ is strongly
suppressed by the bipolar effect due to small level spacing.
However, in the presence of the Coulomb interaction, the effective
level spacing is enlarged and the bipolar effect is weakened,
resulting in $ZT$ to be considerably high. Thus, it is more likely
to find a high efficient thermoelectric QDs with large Coulomb
interaction. By using the parameters for a typical QD, the $ZT$ can
reach over 5.
\end{abstract}

 \maketitle

\section {Introduction} Thermoelectric materials are such materials
that can directly convert the thermal energy into the electrical
energy. Thus the thermoelectric energy conversion technology has
been recognized as the most feasible energy conversion technology.
However, due to its low efficiency, this technology has not been
widely used. Thus it is important to find high efficient
thermoelectric materials. The efficiency of thermoelectrical
materials is measured by the dimensionless thermoelectrical figure
of merit $ZT$, while $ZT=\sigma S^{2}T/\kappa$. Here $S$ is the
Seebeck coefficient, $\sigma$ is the electric conductivity and
$\kappa$ is the total thermal conductivity which contains the
lattice thermal conductivity $\kappa_{l}$ and electric (carrier)
thermal conductivity $\kappa_{e}$, and $T$ is the operating
temperature of the device.\cite{nolas} For a material to be a good
thermoelectric material, it must have a large $ZT$, which means in
order to achieve a large $ZT$, one must increase the Seebeck
coefficient $S$ and electric conductivity $\sigma$ and decrease the
thermal conductivity $\kappa$. However, it seems difficult to have a
high $ZT$ in nature materials. Several reasons hinder the rising of
$ZT$. First, in conventional solids the Wiedemann-Franz law ($
\kappa_{e}/\sigma T=(k_{B} \pi)^{2}/3e^{2} $) is obeyed,\cite{widem}
which means that an increase in the electric conductivity also leads
to an increase in the thermal conductivity. Second, according to the
Mott relation,\cite{mott} an increase in the electric conductivity
is apt to lead to an decrease in Seebeck coefficient. Thus in the
past fifty years, the maximum $ZT$ is holding at about $1$. This
largely affect the industrial applications.

Recently, the advance in nanostructure materials have largely
stimulated the development in thermoelectric materials. Due to the
quantum phenomena emerged in nanostructure materials, the classical
results such as the Mott relation and the Wiedemann-Franz law may
not hold.\cite{kubala} What's more, the thermoelectric properties of
the nanostructure materials can be modulated by changing the gate
voltage. Thus it opens a new and wide road to find efficient
thermoelectric materials. The idea of using low dimensional
structure materials to gain high $ZT$ was first introduced by Hicks
and Dresselhaus in 1993.\cite{dresselhaus} They theoretically show
that $ZT$ increases swiftly as the dimensions decrease, far beyond
the value obtainable in bulk materials. Following this suggestion
and with the development in nanotechnology, various groups were able
to fabricate nanostuctures and measure their thermoelectric
properties.\cite{nanot,hoch, boukai, harman, venk} For example,
Harman \emph{et. al.} have measured the thermoelectric properties of
quantum dot and a maximum value $ZT\approx 2$ was
obtained.\cite{harman} Venkatasubramanian \emph{et. al.} have
measured a thin-film thermoelectric device and have observed a
maximum $ZT$ of $\sim 2.4$ at room-temperature.\cite{venk} Apart
from the experimental efforts, many theoretical studies have been
carried out on low dimensional structures such as quantum dots,
nanowires, and superlattices.\cite{dresselhaus, kim, theory, venk2,
finch} For example, Venkatasubramanian and Chen have concluded that
the main reason of high $ZT$ in low dimensional materials is due to
a significant reduction in lattice thermal conductivity.\cite{venk2}
A giant figure of merit in single-molecule device is obtained by
Finch \emph{et. al.}.\cite{finch} All these efforts show that a high
figure of merit may exist in nanomaterials. However, due to the
complexity and expensiveness of the nanomaterials, there still a
long way to go for the commercial applications of nanostructure
thermoelectric materials. At present the most promising
nanostructure thermoelectric material is nanocomposite
thermoelectric material.\cite{min,widem,pou}

In this paper, we study the thermoelectric properties of
lead-QD-lead system with the QD in the Coulomb blockade regime. The
thermoelectric properties of QD have been widely studied. For
example, Beenakker \emph{et al.} have investigated the thermal
properties of QD with multiple energy levels,\cite{beenakker,kubala}
but they only considered the situation that temperature ($k_B T$) is
much bigger than the level-width ($\Gamma$). On the other hand,
Turek \emph{et al.} and Murphy \emph{et al.} have studied the
thermal properties of the QD in the situation $k_B T \sim
\Gamma$,\cite{turek, murphy} but the QD considered contains only a
single energy level. Here we consider the QD having the multiple
energy levels and in the situation of the temperature smaller than
the interaction $U$. In this regime, some new phenomena emerge. By
using the Landauer-B\"{u}tticker formalism combining the
nonequilibrium Green's functions,\cite{haug} the electronic
conductivity, Seebeck coefficient, and thermal conductivity are
obtained. Due to the electron-hole symmetry, the Seebeck coefficient
is always antisymmetric. If only a single energy level in the QD is
considered, the $ZT$ increases monotonously with temperature $T$,
and $ZT$ can be very large at high temperature, consistent with
previous results. However, with two or multi levels and the
temperature is on the order of energy gap, the $ZT$ is strongly
suppressed by the bipolar effect, mainly caused by the antisymmetric
property of the Seebeck coefficient. On the other hand, when the
Coulomb interaction $U$ is considered, the energy spacings are
enlarged due to the Coulomb blockade effect. The bipolar effect is
greatly reduced and high value of $ZT$ may again be achieved. In a
typical QD, the Coulomb interaction $U$ is usually larger by an
order than the linewidth $\Gamma$ and the single particle energy
spacing. Under these conditions, the $ZT$ can be quite high, with
its maximum value reaching over 5.

\section {Model and formalism} The system of the lead-QD-lead can
be described by the following Hamiltonian:
\begin{eqnarray}\label{a}
 H & = &
 \sum\limits_{\alpha, k, \sigma  }\varepsilon_{\alpha k}\hat{c}_{\alpha k\sigma}^{\dag}\hat{c}_{\alpha
 k\sigma}
 + \sum\limits_{\alpha, k,i, \sigma}t_{\alpha
 k}(\hat{d}_{i \sigma}^{\dag}\hat{c}_{\alpha k\sigma}+H.c.) \nonumber\\
 & + &
 \sum\limits_{i=1,2;\sigma}\varepsilon_{i}
 \hat{n}_{i \sigma} +\frac{U}{2}\sum\limits_{i, \sigma,j, \sigma'(i \sigma\neq j \sigma')}
 \hat{n}_{i \sigma}\hat{n}_{j \sigma'},
\end{eqnarray}
where $\hat{n}_{i \sigma}= \hat{d}_{i \sigma}^{\dagger}\hat{d}_{i
\sigma}$, and $\alpha =L, R$ represent the left and right leads.
$\hat{c}_{\alpha k\sigma}^{\dag}$ and $\hat{d}_{i \sigma}^{\dagger}$
create an electron with spin $\sigma$ in the $\alpha$ lead and the
$i$th energy level of QD, respectively.
Here the intra-dot electron-electron Coulomb interaction is
considered, with the interaction strength $U$. The second term in
Eq. (\ref{a}) describes the tunneling coupling between the QD and
the two leads and $t_{\alpha k}$ is the hopping matrix element.

By using nonequilibrium Green's function methods, the electronic
current and electric thermal current from the left lead flowing into
the QD can be written in the forms:\cite{lee,haug}
\begin{equation}\label{a1}
\left( {\begin{array}{*{20}c}
   I  \\
   Q  \\
\end{array}} \right) = \frac{2}{h}\int {d\omega \left( {\begin{array}{*{20}c}
   { - e}  \\
   {\omega  - \mu _L }  \\
\end{array}} \right)} \left( {f_{L} - f_{R}} \right){\rm T}(\omega
),
\end{equation}
where $f_\alpha=f(\omega-\mu_\alpha)
=1/\{exp[(\omega-\mu_{\alpha})/k_B T]+1\} $ is the Fermi
distribution of the $\alpha$ lead and $\rm T (\omega)$ is the
transmission coefficient. $\rm T (\omega)$ can be expressed by the
following expression:
\begin{equation}\label{a2}
{\rm T}(\omega ) =Tr[ \frac{{\Gamma _L \Gamma _R }}{{\Gamma _L  +
\Gamma _R }} (\mathbf{G}^r  - \mathbf{G}^a )].
\end{equation}
Here $\mathbf{G}^r (\mathbf{G}^a )$ is the standard retarded (advanced) Green¡¯s
function of the QD,\cite{lee,haug} and $\Gamma_{\alpha,ij} =\sum_k
2\pi |t_{\alpha k}|^2 \delta(\omega-\epsilon_{\alpha k})$ is the
linewidth functions which assume to be independent of the energy
$\omega$. We introduce the following integrals $I_n (T)$
($n=0,1,2,..$): $I_n (T) =  - (2/h)\int\limits {\omega ^n (\partial
f/\partial \omega) {\rm T}(\omega)d\omega }$. By using the
quantities $I_n(T)$, the linear electric conductance $G$,
thermopower $S$, and thermal conductance $\kappa$ can be expressed
in very simple forms: \cite{note1}
\begin{subequations}
\renewcommand{\theequation}
{\theparentequation-\arabic{equation}}
\begin{equation}
 G = e^2 I_0 (T),
 \end{equation}
 \begin{equation}
  S =  -
{I_1 (T)}/[{eTI_0 (T)}],
\end{equation}
\begin{equation}
\kappa =(1/T)[ I_2 (T) - I_1^2 (T)/I_0 (T)].
\end{equation}
\end{subequations}
Therefore, the only question left is to calculate the Green's
functions of the QD.

\begin{figure}
\centering
\includegraphics[height=170pt,viewport=39 45 739 583,clip]{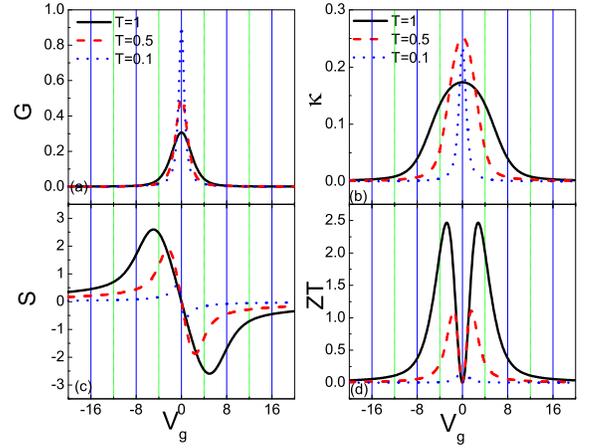}
\caption{ (Color online) $G(2e^{2}/h)$, $k(2k_B/h)$, $S(k_B/e)$, $ZT$ vs.
the level $\epsilon_1$ for the different temperature $T$ for the
single-level QD and $U=0$.}\label{Q1}
\end{figure}

\section {Numerical result} In the numerical investigation, we
consider the symmetric barriers with $\Gamma_{L,ij}= \Gamma_{R,ij}
=\Gamma$, and set $\Gamma=0.5$ and $\mu_R=0$ as the energy zero
point. We consider the linear regime, then $\mu_L
=\mu_R\equiv\mu=0$. First, we study the case with QD possessing only
one energy level $\varepsilon_1$ and in the absence of the Coulomb
interaction ($U=0$). In this case, the Green function of QD can be
easily obtained as $\mathbf{G}_{1\sigma}^{r}(\omega)
=\mathbf{G}_{1\sigma}^{a*}(\omega)=1/(\omega-\varepsilon_{1}+
i\Gamma)$. Inserting this into Eq. (\ref{a2}), the transmission
coefficient ${\rm T(\omega)}$ can be obtained and the thermoelectric
properties can then be calculated straightforwardly. Fig.1 shows the
electric conductance $G$, the thermal conductance $\kappa$, the
thermopower $S$, and $ZT$ versus the level $\varepsilon_1$ for the
different temperature $T$. Variation of $\varepsilon_1$ is
equivalent to variation of the gate voltage in an experimental
setting. The electric conductivity $G$ and the thermal conductance
$\kappa$ exhibit a single resonant peak at the position
$\varepsilon_{1}=0$. The peak height of the thermal conductance
$\kappa$ first increases and then decreases while the increase of
the temperature $k_B T$ (see Fig.1(b)). The reason is as follows.
The thermal conductance is determined by two aspects: the heat
transferred by each electron and the tunneling probability of each
electron. When temperature increases, the average tunneling
probability decreases but the heat transferred by each electron
increases, leading a non-monotonic relation of $\kappa$ and $k_B T$.
The property of thermopower $S$ is described in Fig.1(c). Here we
can see that the curves are antisymmetric due to the electron-hole
symmetry. The reason is as follows. The thermoelectric effect is
caused by the temperature difference. There are more electrons being
excited above the chemical potential $\mu$ in the hotter region and
correspondingly more holes being generated below $\mu$. When the
energy level of QD is below $\mu$, the main carriers are holes and
then the thermal power is positive. When the energy level is above
$\mu$, the main carriers are electrons and thus the thermal power is
negative. So one can adjust the gate voltage or equivalently
$\varepsilon_1$ and obtain the optimized thermal power. Once the
thermal power, the electron conductivity and the thermal
conductivity are known, $ZT$ can be calculated. Fig.1(d) describes
$ZT$ as a function of QD's level $\varepsilon_{1}$. The optimized
$ZT$ can be obtained by modulating $\varepsilon_1$ when the system
is kept at a fixed temperature. With increase of temperature, the
value of optimized $ZT$ also increases and it goes to infinity as
$T$ approaches infinity.This is consistent with the previous
result.\cite{kuo} Of course, this is a non-sensible result due to
only single level being considered here. As we can see, at
temperature $T=1$ (i.e. $T=2\Gamma$), the optimized $ZT$ is about
2.5.

\begin{figure}
\centering
\includegraphics[height=170pt,viewport=44 40 739 583,clip]{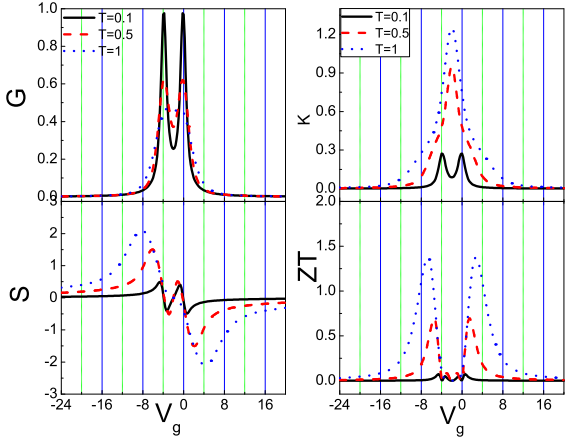}
\caption{ (Color online)  $G(2e^{2}/h)$, $k(2k_B/h)$, $S(k_B/e)$, $ZT$
vs. $V_g$ for the different temperature $T$ for the two-levels QD
and with the parameters $U=0$ and level interval
$\Delta\varepsilon=4$.}\label{Q2}
\end{figure}

In a realistic situation, depending on temperature, dot size, {\it
etc.}, one normally has to consider multi levels. For a multi-level
dot, the spacing between neighboring levels is an important
quantity. For that purpose, investigating a two-level dot will
capture essential physics due to the level spacing. In the following
we study the thermoelectric properties of a QD containing two energy
levels. The Green function of QD with two energy levels
is:
\begin{equation}
\mathbf{G}^r_{\sigma}(\omega) \equiv \left( \begin{array}{ccc}
\mathbf{G}^r_{11\sigma} & \mathbf{G}^r_{12\sigma}\\
\mathbf{G}^r_{21\sigma} & \mathbf{G}^r_{22\sigma}
\end{array} \right) = \mathbf{g}^r_{\sigma}(\omega)
+\mathbf{g}^r_{\sigma}(\omega) \mathbf{\Sigma}^r \mathbf{G}^r_{\sigma}(\omega).
\end{equation}
\begin{equation}
\mathbf{G}^<_{\sigma}(\omega) \equiv \left( \begin{array}{ccc}
\mathbf{G}^<_{11\sigma} & \mathbf{G}^<_{12\sigma}\\
\mathbf{G}^<_{21\sigma} & \mathbf{G}^<_{22\sigma}
\end{array} \right) = \mathbf{G}^r_{\sigma}(\omega) \mathbf{\Sigma}^< \mathbf{G}^a_{\sigma}(\omega).
\end{equation}

Here, the boldface letters ($\mathbf{G}$, $\mathbf{g}$, and
$\mathbf{\Sigma}$) represent the $2 \times 2$ matrix.
$\mathbf{g}^r_{\sigma}$ is the Green Function of QD without coupling
to the leads. $\mathbf{g}^r_{\sigma}$  can be obtained from the
equation of motion technique (the detailed deduction can be seen in
Appendix A) and
 $\mathbf{\Sigma}^{r,<} $ are obtained from Dyson equations ( Here we just consider the
first order of self-energy correction and
have neglected the higher order of self-energy correction that
due to the e-e interaction ): \cite{sun}

\begin{equation}\label{a4}
 \mathbf{g}_{ij\sigma}^r(\omega) = \{ \frac{1 - \{ N_{i\sigma }\} } {\omega - \varepsilon _i  - [N_{i\sigma } ]U } +
 \frac{\{ N_{i\sigma } \} } {\omega - \varepsilon _i  - ([N_{i\sigma } ]+1)U} \} \delta_{ij},
\end{equation}

\begin{equation}
\mathbf{\Sigma}^r_{\sigma} = \left( \begin{array}{ccc}
-i\Gamma & -i\Gamma\\
-i\Gamma & -i\Gamma
\end{array} \right)
\end{equation}

\begin{equation}
\mathbf{\Sigma}^<_{\sigma} = \left( \begin{array}{ccc}
i[\Gamma_L f_L +
\Gamma_R f_R] & 0\\
0 & i[\Gamma_L f_L +
\Gamma_R f_R]
\end{array} \right)
\end{equation}
where $N_{i\sigma}=n_{i\bar \sigma }+n_{\bar{i} \sigma }+n_{\bar{i}\bar
\sigma }$, $ n_{i \sigma}$ is the electron occupation number in the
$i$th energy level with the spin state $\sigma$, $\bar\sigma =
\downarrow$ while $\sigma =\uparrow$ and $\bar\sigma = \uparrow$
while $\sigma =\downarrow$, and $\bar{i}=1$ while $i=2$ and
$\bar{i}=2$ while $i=1$. In Eq.(\ref{a4}), $[N_{i\sigma } ]$ means the
integer part of $N_{i\sigma}$, ${\{ N_{i \sigma } \}
}=N_{i\sigma}-[N_{i \sigma } ]$, namely the decimal part of
$N_{i\sigma}$. In addition, the electron occupation numbers $ n_{i
\sigma} $ need to be self-consistently calculated with the
self-consistent equation $ n_{\sigma} = -i \int (d\omega/2\pi) \mathbf{G}^<_{
\sigma}(\omega)$.

\begin{figure}
\centering
\includegraphics[height=170pt,viewport=40 45 739 573,clip]{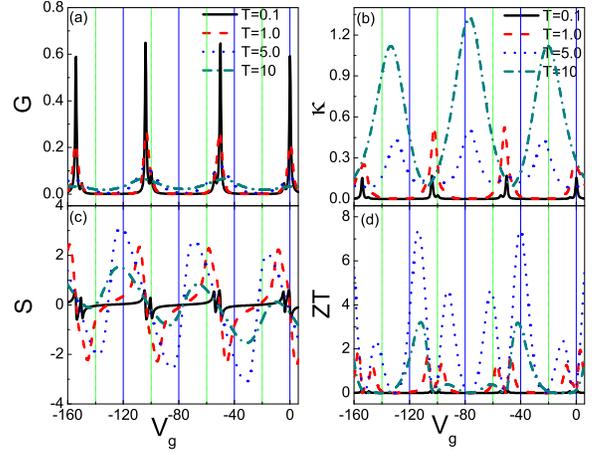}
\caption{ (Color online)  $G(2e^{2}/h)$, $k(2k_B/h)$, $S(k_B/e)$, $ZT$
vs. $V_g$ for the different temperature $T$ for the two-levels QD
with the parameters $U=50$ and level interval
$\Delta\varepsilon=4$.}\label{Q3}
\end{figure}

Fig.2 shows the conductance $G$, the thermal conductivity $\kappa$,
the thermopower $S$, and $ZT$ versus the gate voltage $V_g$ in the
absence of the Coulomb interaction ($U=0$). Here the energy levels
and the gate voltage are related by: $\varepsilon_{1}=V_g$ and
$\varepsilon_{2}=V_g +\Delta \varepsilon$, where $\Delta
\varepsilon$ is the spacing between the two levels. At the low
temperature, $G$ and $\kappa$ in Fig.2(a,b) exhibit two peaks due to
the two energy levels. There seems no great change in $G$ in
comparison with that for the single-level QD. However, behavior of
$\kappa$ in Fig.2(b) is more sensitive to temperature.
When temperature is of the order of $\Delta \varepsilon$,
the peaks are broadened to a degree to give rise to a huge peak.
Meanwhile, the thermopower $S$ and the $ZT$ are largely suppressed
when the lead's chemical potential $\mu$ is between the two energy
levels ($\varepsilon_{2}>\mu>\varepsilon_{1}$). On the other hand,
the optimized $ZT$ remains considerably large when $\mu$ is outside
the two energy levels ($\mu<\varepsilon_{1},\varepsilon_{2} $ or
$\mu>\varepsilon_{1},\varepsilon_{2}$). In a real system there are
many energy levels in a QD. The $ZT$ with the chemical potential
outside of the two levels is influenced by other levels. So we only
focus on the $ZT$ for the case with
$\varepsilon_{2}>\mu>\varepsilon_{1}$, in which the optimized $ZT$
is rather low. This is because the electron and holes are excited in
the range of $k_{B}T$, then the carriers can tunnel through the
$i$th energy level in QD when
$\mu-k_{B}T<\varepsilon_{i}<\mu+k_{B}T$. When temperature is
significantly lower than the level spacing $\Delta \varepsilon$, the
carriers can only choose one energy level to tunnel, which is
similar to the single-level QD. When temperature is on the order of
$\Delta \varepsilon$, the carriers can choose both levels to
tunnel. Due to the temperature difference between the two leads, the
electrons above (below) the chemical potential $\mu$ in the left
lead are more (less) than the electrons in the right lead. While
$\varepsilon_1<\mu<\varepsilon_2$, electrons tunnel from the left
lead to the right lead through the level $\varepsilon_2$, at the
same time electrons tunnel from the right lead to the left lead
through the level $\varepsilon_1$, in other words the hole tunnels
from the left lead to the right lead through $\varepsilon_1$. This
is a bipolar effect: a nonzero heat conduction emerges even when the
net electrical current is zero. Furthermore, the Seebeck coefficient
$S$ is significantly suppressed since the carriers are tunneling
through the QD in both channels via opposite directions. The above
reasons cause $ZT$ to be very small.

From Fig.2(d), we can see that the optimized $ZT$ is arising with
temperature when temperature is much lower than the level spacing
$\Delta \varepsilon$. However, when $T$ is of the order of $\Delta
\varepsilon$, the bipolar effect is enhanced and $ZT$ is decreased.
In order to increase $ZT$, one needs to find a QD with large enough
level spacing $\Delta \varepsilon$ such that to reduce the bipolar
effect. In a typical QD, the bare value of $\Delta \varepsilon$ is
not big enough. An alternative way is to choose a QD with strong
Coulomb interaction $U$. Though the Coulomb interaction may suppress
$ZT$,\cite{kuo} at the same time it can broaden the energy level.
The effective level spacings can be widened to $U+\Delta
\varepsilon$, which can give rise to a large $ZT$. In a typical QD,
$U$ can be quite large. For example, $U$ in $C_{60}$ is on the order
of 0.1 $ev$.\cite{park} In addition, the region that reduces the
bipolar effect is $U>k_{B}T/10$. In this situation, the suppression
of $ZT$ due to Coulomb interaction can be neglected.

\begin{figure}
\centering
\includegraphics[height=150pt,viewport=50 50 739 573,clip]{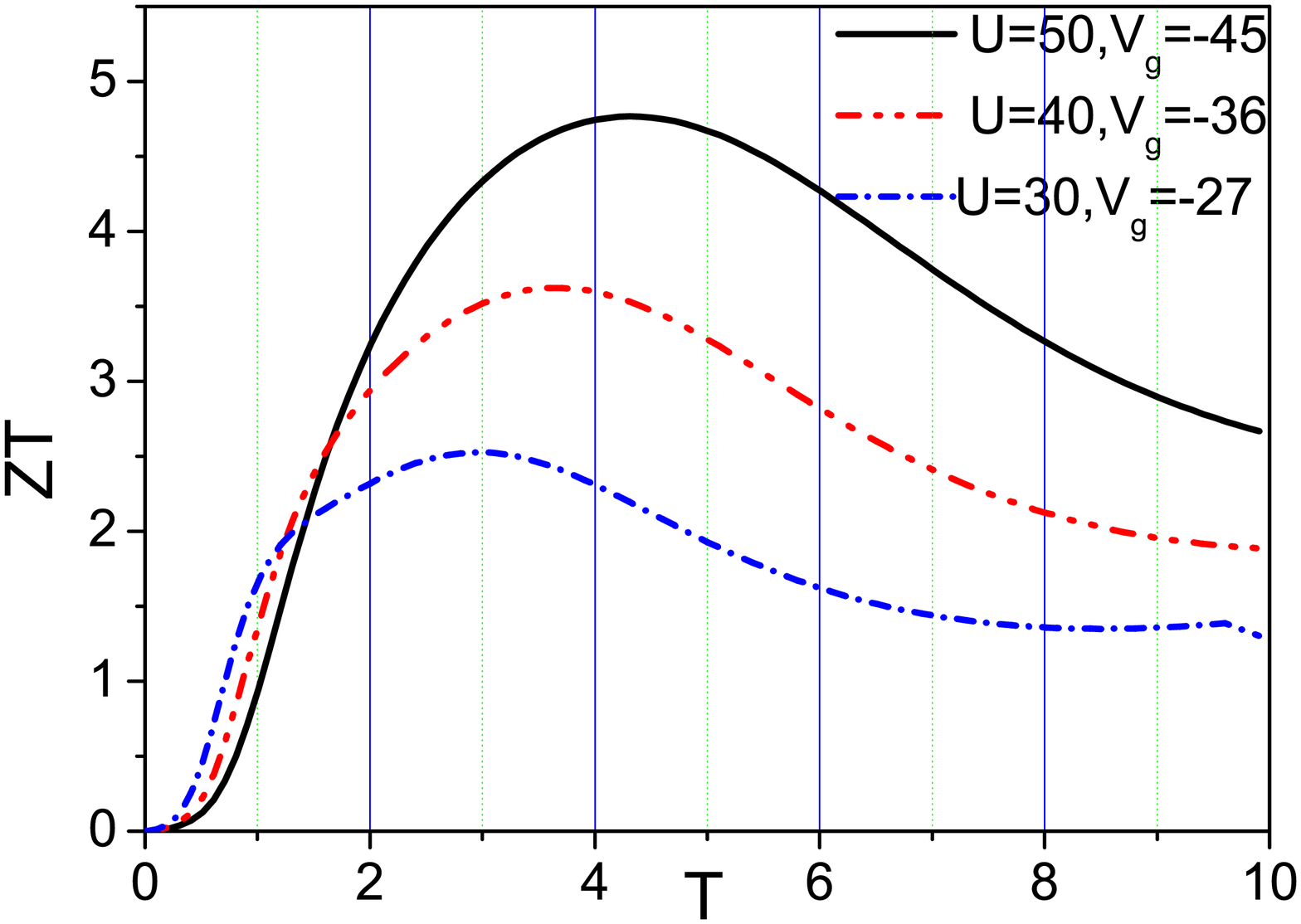}
\caption{ (Color online) $ZT$ vs. temperature $T$ for the different
coulomb interaction $U$ with $\Delta\varepsilon=4$. For each
situation we choose a suitable gate voltage.}\label{Q4}
\end{figure}

  Now we consider that the QD contains two energy levels and both
inter-level and intra-level Coulomb interactions exist. For
convenience, we set the inter-level interaction to be equal to
intra-level interaction. The interaction $U$ is about one order
larger than the level spacing $\Delta\varepsilon$, as for a typical
QD. In the existence of the interaction $U$, the levels are located
at $\varepsilon_1$, $\varepsilon_1+U$,
$\varepsilon_1+\Delta\varepsilon +2U$, and
$\varepsilon_1+\Delta\varepsilon +3U$ due to the Coulomb blockade
effect.Thus the linear electric conductivity $G$ exhibits four main
resonant peaks at the positions of $V_g=0$, $-U$,
$-2U-\Delta\varepsilon$ and $-3U-\Delta\varepsilon$ (see Fig.3a). In
addition, there are also some smaller peaks at low temperature due
to the tunneling through the excite states.
Fig.3(b) shows the thermal conductivity $\kappa$ versus the gate
voltage. We can see that at high temperature ($k_BT
\sim U$) the bipolar effect occurs and huge peaks of the thermal
conductivity emerge at the valleys between the two main adjacent
peaks of the conductivity.

  The thermopower $S$ shown in Fig.3(c) is
sensitive to the slope of conductivity.
It is clearly seen that at low temperature the thermal power changes
from positive to negative when the gate voltage moves across each
peak of the conductivity (e.g. T=0.1).
With temperature rising, the small peaks in the conductivity are
absorbed to the main peaks and accordingly the transitions from
positive to negative of the thermal power are reduced to four. The
peak and valley values of the thermal power are enhanced with rising
temperature. However, due to the bipolar effect, the peaks and
valleys reach their maximum values at about $T=5$ (i.e. $U/10$).
Further increasing of $T$ will decrease these values. After $G$,
$\kappa$, and $S$ are calculated, the $ZT$ can be determined, shown
in Fig.3(d). Since the Coulomb interaction $U$ broadens the level
spacing, the bipolar effect is greatly suppressed and $ZT$ is
enhanced. The optimized $ZT$ can be over $5$, much larger than the
value in Fig.2 without interaction.

  To further investigate the effect of temperature and Coulomb
interaction, we numerically calculate the $ZT$ versus temperature at
different Coulomb interaction $U$ (see Fig.4). Notice that the
bipolar effect can be enhanced with increasing temperature but
weakened with broadening of the energy spacing. Thus when $k_B
T<U/10$, $ZT$ is enhanced with rising temperature because of the
weak bipolar effect. When $k_B T>U/10$, $ZT$ saturates with further
increasing of temperature. Moreover, with increase of $U$, the
optimized $ZT$ also increases.

\section {Conclusion} We investigate the thermoelectric properties
of a QD that contains one or two levels and is in the Coulomb
blockade regime. The results exhibit that in the absence of the
Coulomb interaction, the $ZT$ can be very high if only one level in
the QD is considered, but the $ZT$ is greatly suppressed with multi
levels due to the bipolar effect. When the Coulomb interaction $U$
is considered in the QD, the spacings of energy levels are
increased, and the bipolar effect is weakened, thus the $ZT$ can be
considerably high. For an actual QD in which its Coulomb interaction
is one order larger than the level spacing, the optimized $ZT$ can
be over $5$, much larger than the values from natural materials.

\section {Acknowledge} We gratefully acknowledge the financial
support from the Chinese Academy of Sciences, NSF-China under Grants
Nos. 10525418, 10734110, and 10821403 and National Basic Research
Program of China (973 Program project No. 2009CB929100). X.C.X. is
supported by US-DOE under Grants No. DE-FG02- 04ER46124 and C-SPIN
center in Oklahoma.

\begin{appendix}
\section {The Green Function of QD}

In this appendix we give a detailed deduction on how to get the
Green function of the QD with two energy levels.

First, a single QD without coupling to the leads can be described by
the following Hamiltonian:
\begin{equation}
H_{D} = \sum\limits_{i=1,2;\sigma}\varepsilon_{i}
 \hat{n}_{i \sigma} +\frac{U}{2}\sum\limits_{i, \sigma,j, \sigma'(i \sigma\neq j \sigma')}
 \hat{n}_{i \sigma}\hat{n}_{j \sigma'}
\end{equation}
For an isolated QD, the exact Green function can be obtained by the
equation of motion technique:
\begin{eqnarray}
 \mathbf{g}_{ij\sigma }^r  &=& \left\{\frac{{(\omega  - \varepsilon _i  - U) +
U(\left\langle {n_{i\bar \sigma } } \right\rangle  + \left\langle {n_{i'\sigma } } \right\rangle
+ \left\langle {n_{i'\bar \sigma } } \right\rangle )}}{{(\omega  - \varepsilon _i )(\omega  - \varepsilon _i  - U)}}
 \right.
 \nonumber \\
   &+& \frac{{2U^2 (\left\langle {n_{i\bar \sigma } n_{i'\sigma } } \right\rangle  + \left\langle {n_{i'\sigma } n_{i'\bar \sigma } }
\right\rangle  + \left\langle {n_{i\bar \sigma } n_{i'\bar \sigma } } \right\rangle )}}
{{(\omega  - \varepsilon _i )(\omega  - \varepsilon _i  - U)(\omega  - \varepsilon _i  - 2U)}} \nonumber \\
 &+& \left. \frac{6U^3 \left\langle n_{i\bar \sigma } n_{i'\sigma }
n_{i'\bar \sigma }  \right\rangle } {(\omega  - \varepsilon _i
)(\omega  - \varepsilon _i  - U)(\omega  - \varepsilon _i  - 2U)
(\omega  - \varepsilon _i  - 3U) } \right\}\delta_{ij} .
\end{eqnarray}
This is an exact solution without any approximation. However, it is
difficult to self-consistently calculate $\langle n_{i \sigma }
n_{i'\sigma' } \rangle$ and $\left\langle {n_{i\bar \sigma }
n_{i'\sigma } n_{i'\bar \sigma } } \right\rangle$ through numerical
means, and some approximations are needed. Here, we make the
approximation $\langle n_{i \sigma } n_{i'\sigma' } \rangle = 0 $
while $\langle n_{i \sigma } \rangle +\langle n_{i'\sigma' }
\rangle<1$ and $\langle n_{i \sigma } n_{i'\sigma' } \rangle =
\langle n_{i \sigma } \rangle +\langle n_{i'\sigma' } \rangle-1 $
while $\langle n_{i \sigma } \rangle +\langle n_{i'\sigma' }
\rangle>1$. In addition, we make another approximation $\left\langle
{n_{i\bar \sigma } n_{i'\sigma } n_{i'\bar \sigma } } \right\rangle
=0$ while $N_{i \sigma}<2$ and $\left\langle {n_{i \bar{\sigma} }
n_{i'\sigma } n_{i'\bar \sigma } } \right\rangle =\{N_{i \sigma}\}$
while $N_{i \sigma}>2$. These approximations are reasonable since
the fluctuation of the occupation number in the QD is less than one
at zero bias and the temperature $k_B T <U$.\cite{sun} Thus, when
$N_{i \sigma}<1$, the Green function can be simplified as:
\begin{equation}
\mathbf{g}_{ij\sigma }^r  = \{\frac{{1 - \{ N_{i\sigma } \} }}{{\omega  - \varepsilon _i }} + \frac{{\{ N_{i\sigma } \} }}{{\omega  - \varepsilon _i  - U}}\}\delta_{ij};
\end{equation}
When $1<N_{i \sigma}<2$, the Green function can be simplified as:
\begin{equation}
\mathbf{g}_{ij\sigma }^r  = \{\frac{{1 - \{ N_{i\sigma } \} }}{{\omega  - \varepsilon _i  - U}} + \frac{{\{ N_{i\sigma } \} }}{{\omega  - \varepsilon _i  - 2U}}\}\delta_{ij};
\end{equation}
When $2<N_{i \sigma}<3$, the Green function can be simplified as:
\begin{equation}
\mathbf{g}_{ij\sigma }^r  = \{\frac{{1 - \{ N_{i\sigma } \} }}{{\omega  - \varepsilon _i  - 2U}} + \frac{{\{ N_{i\sigma } \} }}{{\omega  - \varepsilon _i  - 3U}}\}\delta_{ij};
\end{equation}
Then the final form of Green function can be written as:
\begin{equation}
 \mathbf{g}_{ij\sigma}^r(\omega) = \{ \frac{1 - \{ N_{i\sigma }\} } {\omega - \varepsilon _i  - [N_{i\sigma } ]U } +
 \frac{\{ N_{i\sigma } \} } {\omega - \varepsilon _i  - ([N_{i\sigma } ]+1)U} \} \delta_{ij},
\end{equation}

\end{appendix}


\begin{thebibliography}{99}
\bibitem{note1} Here we have neglected the lattice thermal conductance since
it is usually a constant in a given
material, normally small in a low dimensional device.
\end{thebibliography}
\end{document}